\documentclass[twocolumn,showpacs,pra,aps]{revtex4}

\usepackage{graphicx}
\usepackage{bm}
\usepackage{psfrag}
\usepackage{amsmath}
\usepackage{amssymb}
\usepackage{latexsym}
\usepackage{exscale}
\usepackage{natbib}

\newcommand{\ten}[1]{\mbox{\textbf{\textsf{#1}}}}
\begin{document}

\title{Van der Waals interaction and spontaneous decay of an excited
atom in a superlens-type geometry}

\author{Agnes Sambale}
\author{Dirk--Gunnar Welsch}

\affiliation{Theoretisch--Physikalisches Institut,
Friedrich--Schiller--Universit\"at Jena,
Max--Wien-Platz 1, D-07743 Jena, Germany}

\author{Ho Trung Dung}
\affiliation{Institute of Physics, Academy of
Sciences and Technology, 1 Mac Dinh Chi Street,
District 1, Ho Chi Minh city, Vietnam}

\author{Stefan Yoshi Buhmann}
\affiliation{Quantum Optics and Laser Science, Blackett Laboratory,
Imperial College London, Prince Consort Road,
London SW7 2BW, United Kingdom}

\date{\today}

\begin{abstract}

Within the framework of macroscopic quantum electrodynamics, the resonant
van der Waals potential experienced by an excited two-level atom near
a planar magneto-electric two-layer system consisting of a slab of
left-handed material and a perfect mirror is studied. It is shown that
disregarding of material absorption leads to unphysical results,
with divergent values for the potential away from the surface. Under
appropriate conditions, the setup is found to feature a barrier near
the surface which can be employed to levitate particles or used as a
trapping or cooling mechanism. Finally, the problem of spontaneous
decay [J. K\"{a}stel and M. Fleischhauer, Phys.~Rev.~A~\textbf{68},
011804(R) (2005)] is revisited. Disregarding of absorption is shown to
drastically falsify the dependence on the atomic position of the decay
rate.
\end{abstract}

\pacs{12.20.-m, 42.50.Wk, 34.35.+a, 42.50.Nn}

\maketitle


\section{Introduction}
\label{sec1}

The first most comprehensive account of the properties of left-handed
materials (LHMs), i.e., materials simultaneously exhibiting negative
real parts of the electric permittivity and the magnetic permeability
in a certain frequency interval, was given by
Veselago~\cite{Veselago1968}. As he pointed out, a negative refractive
index may be attributed to such materials, leading to a number of
unusual optical phenomena. While natural materials with a negative
real part of the permittivity such as plasmas~\cite{Gurnett2006} or a
negative real part of the permeability such as some
ferrites~\cite{Mills1974} exist, LHMs have not yet been found in
nature. During the last decade, significant progress in the
fabrication of LHMs in laboratories has been achieved. Metamaterials
based on periodic arrays of split ring resonators and a grid of
metallic rods have been reported to exhibit left-handedness in the
microwave region~\cite{Smith2000,Shelby2001}. More recently, a
two-dimensional material consisting of a metal--insulator--metal
waveguide has been experimentally demonstrated to show a negative
refractive index in the optical region~\cite{Lezec2007}.
The effective permittivity and permeability of such metamaterials,
valid on length scales which are sufficiently large in comparison to
the elementary building blocks of the material, are determined either
theoretically \cite{Pendry2000, Vinogradov2008} or by means of
reflection experiments \cite{Paul2008}. 

As already indicated in Ref.~\cite{Veselago1968}, a planar
LHM slab with a negative refractive index
has the interesting property of being able to
focus light---an effect which suggests the possibility of realizing
a so-called superlens for subwavelength imaging with
a resolution well beyond the diffraction limit~\cite{Pendry2000}.
The superlens concept has been a subject
of intense discussion~\cite{Kik2004,Ramakrishna2003}, and limiting
factors such as the finite dimension of the lens~\cite{Chen2006} or
the influence of absorption have been studied~\cite{Smith2003}.

In view of the unusual properties of LHMs, it has also been
of interest to study the modifications of dispersion forces in
the presence of magneto-electric materials. In particular, the
question of whether a left-handedness may lead to a repulsive Casimir
force between two magneto-electric plates has been
addressed~\cite{Henkel2005}. From the study of van der Waals (vdW)
forces on single~\cite{Kampf2005} and between two ground-state
atoms~\cite{Spagnolo2007} in the presence of a planar system
exhibiting magneto-electric properties, it has been shown that the
attractive/repulsive behavior of the potential is influenced by the
strength of the magnetic properties rather than by the
left-handedness. This may be attributed to the fact that in both cases
the atoms are in the ground state, so the vdW forces represent
off-resonant interactions that involve integrals over the whole
frequency range. Hence, frequency intervals where no left-handedness
occurs will inevitably also contribute to the force and counteract the
effect of the left-handedness; this becomes evident after expressing
the force in terms of an integral over the imaginary frequency axis
where both the permittivity and the permeability are always real and
positive. Similar arguments also apply to the abovementioned Casimir
force between two magneto-electric plates, where a repulsive behavior
has also been found to primarily require strong magnetic rather than
left-handed properties~\cite{Henkel2005}.

Two strategies might be envisaged to enhance the influence of
left-handed properties on dispersion forces. The first one would be
to consider a material exhibiting left-handedness over a sufficiently
large (real) frequency range, as recently discussed in the context of
the Casimir force~\cite{Leonhardt2007}; however, such an requirement
is, in general, in conflict with the causality requirements met by all
physical media. Secondly, one may consider excited systems like
amplifying media as also addressed in Ref.~\cite{Leonhardt2007}. The
fundamental difference between absorbing and amplifying media
requires, however, a careful reconsideration of the underlying quantum
electrodynamics~\cite{Raabe2008}. The situation is more transparent in
the case of an excited atom where the necessary theory of vdW forces
is readily available~\cite{Buhmann2004}. Excited atoms give rise to
resonant force components which directly depend on the frequencies of
possible downward transitions. If a material exhibits left-handed
properties at one such frequency, one may expect these to influence
the dispersion forces in a more significant way.

The spontaneous emission of an excited atom situated in vacuum
in the vicinity of a perfect mirror combined with a lossless LHM slab
is studied in Refs.~\cite{Fleischhauer2005,Fleischhauer2005a}, where
it is noticed that a slab of unity negative refraction \mbox{($n$
$\!=$ $\!-1$)} has the peculiar feature of making the space between
atom and mirror appear of zero optical length when the atom--slab
distance is equal to the thickness of the slab. For such positions,
called focal points, complete suppression of the spontaneous emission
for a dipole moment parallel to the surface, and enhancement of the
spontaneous emission rate by a factor of two for a dipole moment
perpendicular to the surface is reported. More recently, it has been
pointed out that in reality inhibition of spontaneous decay can be
weakened due to nonradiative decay at short distances and due to
radiative decay at large distances~\cite{Xu07}.

Motivated by Refs.~\cite{Fleischhauer2005, Fleischhauer2005a},
we shall study the vdW interaction of an excited two-level atom with
the quantized electromagnetic field in the presence of a perfect
mirror combined with a realistic LHM slab. To conform with causality
and to avoid unphysical consequences, absorption in the LHM is taken
into account from the very beginning. The paper is organized as
follows. In Sec.~\ref{Sec2}, basic formulas for the vdW potential and
the decay rate, and for the Green tensor of a planar magneto-electric
multilayer structure are given. Section~\ref{sec3} is devoted to
analytical and numerical studies of the vdW potential experienced by
an  excited two-level atom in front of a system consisting of a LHM
slab combined with a perfect mirror. Special emphasis is put on the
effects of material absorption. In Sec.~\ref{spe}, the atomic
spontaneous decay rate is discussed and a comparison with earlier
results is made. Some concluding remarks and a summary are given in
Sec.~\ref{sum}.


\section{Basic equations}
\label{Sec2}

Consider the vdW potential
of a neutral, unpolarized and nonmagnetic atomic system [position
$\mathbf{r}_A$, energy eigenstate
$|n\rangle$, transition frequencies $\omega_{nk}$,
electric-dipole transition matrix elements $\mathbf{d}_{nk}$,
polarizability $\bm{\alpha}(\omega)$] in the presence of an
arbitrary magneto-electric medium of complex permittivity
$\varepsilon(\mathbf{r},\omega)$ and permeability
$\mu(\mathbf{r},\omega)$. The medium affects
the classical Green tensor $\ten{G}(\mathbf{r},\mathbf{r}',\omega)$ of
the electromagnetic field, which satisfies the differential equation
\begin{equation}
\label{DGLgreen}
\left[\bm{\nabla}\times\frac{1}{\mu(\mathbf{r},\omega)}\bm{\nabla}
\times\;-\frac{\omega^2}{c^2}\varepsilon(\mathbf{r},\omega)\right]
\ten{G}(\mathbf{r},\mathbf{r'},\omega)=\bm{\delta}(\mathbf{r}-\mathbf{r'})
\end{equation}
together with the boundary condition at infinity
\begin{equation}
\ten{G}(\mathbf{r},\mathbf{r'},\omega)\rightarrow 0
\quad\mathrm{for}\quad
\left|\mathbf{r}-\mathbf{r'}\right|\rightarrow \infty.
\end{equation}
The potential can be decomposed into a resonant and an off-resonant part
(see, e.g., Ref.~\cite{Buhmann2004}),
\begin{equation}
\label{equ0}
U_n(\mathbf{r}_A)
=U_n^{\mathrm{r}}(\mathbf{r}_A)+U_n^{\mathrm{or}}(\mathbf{r}_A),
\end{equation}
where
\begin{align}
\label{equ1}
&U_n^{\mathrm{r}}(\mathbf{r}_A)=
\nonumber\\
&-\mu_0\sum _k \Theta(\omega_{nk})\omega_{nk}^2\mathbf{d}_{nk}\cdot
\operatorname{Re}\ten{G}^{(1)}(\mathbf{r}_A,\mathbf{r}_A,\omega_{nk})
\cdot\mathbf{d}_{kn}
\end{align}
($\Theta$: unit step function) and
\begin{equation}
\label{equ2}
U_n^{\mathrm{or}}(\mathbf{r}_A)=\frac{\hbar \mu_0}{2\pi}
\int _0^{\infty}\mathrm{d}u\, u^2 \operatorname{Tr}\!
\left[\bm{\alpha}
(iu)\!\cdot\!
\ten{G}^{(1)}(\mathbf{r}_A,\mathbf{r}_A,iu)\right],
\end{equation}
respectively. Here, $\ten{G}^{(1)}(\mathbf{r},\mathbf{r}',\omega)$
refers to the scattering part of the Green tensor. For an excited
atom, the resonant part of the potential often dominates and the
off-resonant part can be neglected,
as we will do throughout this paper.
We shall also be interested in the
rate of the spontaneous decay $|n\rangle\to|k\rangle$ of the atom,
which reads as (see, e.g., Ref.~\cite{Dung2003a})
\begin{equation}
\label{gamma}
\Gamma =
\Theta(\omega_{nk})
\frac{2\omega_{nk}^2}{\hbar \varepsilon_0c^2}\,
{\bf d}_{nk}\cdot {\rm Im}\,
\ten{G}(\mathbf{r}_A,\mathbf{r}_A,\omega_{nk})\cdot
{\bf d}_{nk}.
\end{equation}

\begin{figure}[t]
\begin{center}
\includegraphics[width=\linewidth]{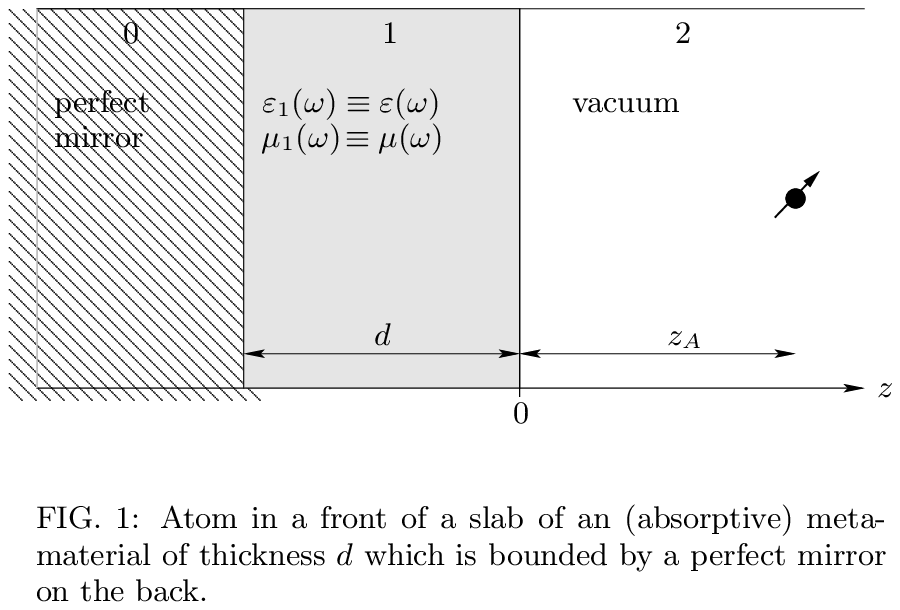}
\end{center}
\caption{Atom in a front of
a slab of an (absorptive)
meta-material of thickness $d$
which is bounded by a perfect mirror on the
back.
}
\label{lhmfig}
\end{figure}

In what follows, we restrict ourselves to a two-level atom
with upper level $|1\rangle$ and lower level $|0\rangle$.
Consider the atom to be placed in free space in front of a
meta-material slab of thickness $d$, permittivity
$\varepsilon(\omega)$ and permeability
$\mu(\omega)$ which is bounded by a perfectly conducting mirror
on the back, as sketched in Fig.~\ref{lhmfig}. The
coordinate system is chosen such that the $z$-axis is perpendicular
to the slab and its origin coincides with the slab--vacuum interface.
The scattering part
of the Green tensor at the relevant atomic transition frequency and
equal positions ${\bf r}\!=\!{\bf r}'\!=\!\mathbf{r}_A$ in
the vacuum region is given by~\cite{Kampf2005}
\begin{multline}
\label{equ14}
\ten{G}^{(1)}(\mathbf{r}_A,\mathbf{r}_A,\omega_{nk})\equiv
\ten{G}^{(1)}(z_A,z_A,\omega_{10})
\\
=\frac{i}{8\pi^2}\int \mathrm{d}^2q\,\frac{1}{\beta}
\sum _{\sigma=s,p}\mathbf{e}_\sigma^+\mathbf{e}_\sigma^-
r_{2-}^{\sigma}e^{2i\beta z_A}.
\end{multline}
In Eq.~(\ref{equ14}), $p$ ($s$) denotes $p$- ($s$-)polarization. The
reflection coefficients are given by
\begin{align}
\label{r2-s}
r_{2-}^s=& \frac{r_{21}^s-
e^{2i\beta_1d}}{1-r^s_{21}
e^{2i\beta_1d}}\,,\\
\label{r2-p}
r_{2-}^p=& \frac{r_{21}^p+e^{2i\beta_1d}}
 {1+r^p_{21}e^{2i\beta_1d}}\,,
\end{align}
with
\begin{equation}
\label{r21}
r^s_{21}=\frac{\mu
\beta-\beta_{1}}{\mu
\beta+
\beta_{1}}\,,\quad
r^p_{21}=\frac{\varepsilon
\beta-
\beta_{1}}
{\varepsilon
\beta+
\beta_{1}}\,,
\end{equation}
where the wave vectors in the $z$ direction are given by
\begin{equation}
\label{equ6b}
\beta=\sqrt{\frac{\omega_{10}
^2}{c^2}-q^2}\,,\quad
\beta_1= \sqrt{k_1^2-q^2}\,,
\end{equation}
with $k_1$, which always appears in the form of $k_1^2$ in the Green
tensor, Eq.~(\ref{equ14}), being given according to
\begin{equation}
\label{equ6}
k_1^2=\frac{\omega_{10}^2}{c^2}\varepsilon
(\omega_{10})\mu(\omega_{10}).
\end{equation}
The $s$- and $p$-polarization unit vectors are defined as
\begin{equation}
\label{equ5}
\mathbf{e}_s^\pm=\mathbf{e}_q\times\mathbf{e}_z,\quad
\mathbf{e}_p^\pm=\frac{c}{\omega_{10}}
(q\mathbf{e}_z\mp\beta\mathbf{e}_q)
\end{equation}
($\mathbf{e}_q=\mathbf{q}/q$, $q=|\mathbf{q}|$). The square root in
Eq.~(\ref{equ6b}) has to be chosen such that it obeys the physical
requirement $\operatorname{Im}\beta_1>0$ for a passive medium, with
this choice being unique for purely absorbing media
($\operatorname{Im}\varepsilon,\operatorname{Im}\mu>0$)
\cite{Vinogradov2008}. The imaginary part of $\beta_1^2$ reads
Im\,$\beta_1^2$ = (Re\,$\varepsilon$\,Im\,$\mu$
+ Im\,$\varepsilon$\,Re\,$\mu$)$\omega_{10}^2/c^2$.
It is positive for right-handed materials, which
together with the condition $\operatorname{Im}\beta_1>0$
implies that $\beta_1$ has to be in the first quadrant of
the complex plane. On the other hand, Im\,$\beta_1^2$
is negative for left-handed materials.
This, combined with the condition $\operatorname{Im}\beta_1>0$,
leads to that $\beta_1$ lies in the second quadrant of 
the complex plane (see also Ref.~\cite{Ramakrishna2005}).
%

Introduction of polar coordinates in the $q$-plane, i.\,e.,
\mbox{$\mathbf{e}_q\!=\!(\cos \phi, \sin \phi,0)$},
\mbox{$\mathbf{e}_s^\pm\!=\!(\sin \phi, -\cos \phi, 0)$},
\mbox{$\mathbf{e}_z\!=\!(0,0,1)$}, implies the identities
\begin{equation}
\mathbf{e}_s^+\mathbf{e}_s^-=
\left(
\begin{array}{ccc}\sin^2 \phi & -\sin \phi \cos \phi & 0 \\
-\sin \phi \cos \phi & \cos^2 \phi & 0 \\
0 &0 & 0
\end{array}
\right)
\end{equation}
and, with $\mathbf{e}_p^\pm=(\mp \beta c \cos \phi /\omega_{10}, \mp
\beta c \sin \phi /\omega_{10}, qc/\omega_{10})$,
\begin{equation}
\mathbf{e}_p^+\mathbf{e}_p^-\!=\!\frac{c^2}{\omega_{10}^2}
\left(\!
\begin{array}{ccc}\!-\beta^2\cos^2 \phi\! &
\!-\beta^2\sin \phi \cos \phi \! & \!-q\beta  \cos \phi\!\\
\!-\beta^2\sin \phi \cos \phi \! & \!-\beta^2 \sin^2 \phi\!
& \!-q\beta  \cos \phi \!\\
\!q\beta \cos \phi \! & \!q\beta \sin \phi \! & \!q^2\!
\end{array}
\!\right)\!.
\end{equation}
With $\mathrm{d}^2q=q\,\mathrm{d}q\,\mathrm{d}\phi$, the angular
integration in Eq.~(\ref{equ14}) can be performed on using
\begin{equation}
\int _0 ^{2\pi} \mathrm{d}\phi\, \mathbf{e}_s^+\mathbf{e}_s^-=\pi
\left(
\begin{array}{ccc}
1&0&0\\
0&1&0\\
0&0&0
\end{array}
\right),
\end{equation}
\begin{equation}
\int _0 ^{2\pi}\mathrm{d}\phi\,
\mathbf{e}_p^+\mathbf{e}_p^-=\frac{\pi c^2}{\omega_{10}^2}
\left(\begin{array}{ccc}-\beta^2 & 0& 0 \\
0& -\beta^2 &0 \\
0 &0 & 2q^2
\end{array}
\right),
\end{equation}
resulting in
\begin{multline}
\label{equ22}
\ten{G}^{(1)}(z_A,z_A,\omega_{10})
=\frac{i}{8\pi}\int _0^ \infty \mathrm{d}q\, \frac{q}{\beta}\,
e^{2i\beta z_A}\\
\times\left(
\begin{array}{c c c}
r^s_{2-}-\frac{\beta^2c^2}{\omega_{10}^2} r^p_{2-} & 0& 0 \\
0& r^s_{2-}-\frac{\beta^2c^2}{\omega_{10}^2} r^p_{2-} &0 \\
0 &0 & \frac{2 q^2 c^2}{\omega_{10}^2} r^p_{2-}
\end{array}
\right).
\end{multline}
From Eq.~(\ref{equ22}) it can be seen that an atom with a dipole
moment perpendicular to the surface is coupled to the $p$-polarized waves
only while an atom with a dipole moment parallel to the surface is
coupled to both $p$- and $s$-polarized waves.
It is instructive to decompose the integral in Eq.~(\ref{equ22})
into two parts,
\begin{multline}
\label{equ22.1}
    \int_0^\infty \!\mathrm{d}q\,
    \frac{q}{\beta}\, e^{2i\beta z_A} f(q)
    \!=\! \int_0 ^{\frac{\omega_{10}}{c}} \!\mathrm{d}\beta\,
    e^{2i\beta z_A} f\!\left(\sqrt{\frac{\omega_{10}^2}{c^2}-\beta^2}\right)
\\
    +\frac{1}{i}\int_0 ^{\infty}\mathrm{d}b
     e^{-2b z_A} f\left(\sqrt{\frac{\omega_{10}^2}{c^2}+b^2}\right).
\end{multline}
The first integral, which contains an oscillating factor,
results from propagating waves, whereas the second one, which
contains an exponentially decaying factor,
results from evanescent waves. The first is the more important
one away from the surfaces while the second dominates near
the surfaces.


\section{vdW potential}
\label{sec3}


\subsection{Ideal (nonabsorbing) LHM}
\label{sec3.1}

Let us assume hypothetically an absolutely nonabsorbing
LHM having \mbox{$\varepsilon
\!=\!\mu\!=\!-1$}.
In accordance with the remarks below Eq.~(\ref{equ5}), we thus set
\begin{equation}
\label{beta_1}
\beta_1=\begin{cases}-\beta\quad\mbox{for }q\le\omega/c,\\
\hspace{1.75ex}\beta\quad\mbox{for }q\ge\omega/c,
\end{cases}
\end{equation}
and the reflection coefficients (\ref{r2-s})--(\ref{r21}) become
\begin{equation}
\label{equ13}
r_{2-}^s=-e^{-2i\beta d}, \quad r_{2-}^p=e^{-2i\beta d}.
\end{equation}
Note that this result does not depend to the sign of the
square root chosen for $\beta_1$, since for the ideal LHM discussed
here, the reflection coefficients are invariant under a change
$\beta_1\to -\beta_1$.
Substitution of these reflection coefficients in
Eq.~(\ref{equ22}) for the Green tensor leads to
\begin{multline}
\label{gtensor}
   \ten{G}^{(1)}(z_A,z_A,\omega_{10})=-\frac{i}{8\pi}
     \int^{\omega_{10}/c}_0 \!\mathrm{d}\beta\;
     e^{2i\beta(z_A-d)}
\\
\times\left(
\begin{array}{c c c}
1+\beta^2c^2/\omega_{10}^2 & 0& 0 \\
0& 1+\beta^2c^2/\omega_{10}^2 &0 \\
0 &0 & -2 (1- c^2\beta^2/\omega_{10}^2)
\end{array}
\right)
\\
- \frac{1}{8\pi}
   \int _0^{\infty}\!\mathrm{d}b \;
e^{-2b(z_A-d)}
\\
\times\left(
\begin{array}{c c c}
1-b^2c^2/\omega_{10}^2 & 0& 0 \\
0& 1-b^2c^2/\omega_{10}^2 &0 \\
0 &0 & -2 (1+ c^2b^2/\omega_{10}^2)
\end{array}
\right).
\end{multline}

\begin{figure}[t]
\begin{center}
\includegraphics[width=\linewidth]{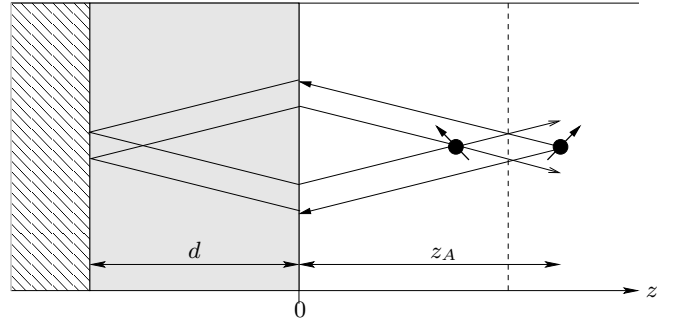}
\end{center}
\caption{Image-dipole construction for the setup depicted in
Fig.~\ref{lhmfig}. The dashed line marks the position of a perfect
mirror that would generate the same image.}
\label{imfig}
\end{figure}

Equation (\ref{gtensor}) reveals that for
$z_A>d$, the Green tensor exactly coincides
with that of a configuration in which a perfectly conducting mirror is placed at
$z=d$. As can be seen from an image dipole construction, this is due to the
perfect negative refraction taking place at the vacuum--LHM
interface: The image of an electric dipole at $z_A>d$ is situated at
$z_A^\star=d-(z_A-d)$, just as it would be if a perfectly conducting
mirror were placed at $z=d$ (see Fig.~\ref{imfig}).
%
An important difference to the case of a mirror placed at $z=d$ is the
fact that in the setup considered here the atom can be placed
in the region $0<z_A\le d$. However, it
can be seen that, whereas the first integral in Eq.~(\ref{gtensor}),
which is complex, is always well-behaved,
the second (purely real) integral tends to minus
infinity in this region. Thus we are left with
an infinite potential for $0 < z_A \leq d$,
despite the absence of any physical surface at $d$.
We clearly have a situation where an unphysical assumption
(of absolute zero absorption) gives rise to unphysical results.

After calculating
the two integrals, we obtain for \mbox{$z_A>d$}
\begin{align}
\label{gr1}
G_{xx}^{(1)}(z_A,z_A,\omega_{10})&=G_{yy}^{(1)}(z_A,z_A,\omega_{10})
\nonumber\\
&=\frac{\omega_{10} e^{i\tilde{z}}}{4\pi c\tilde{z}^3}
\left(1-i\tilde{z}-\tilde{z}^2\right),
\end{align}
\begin{equation}
\label{gr2}
G_{zz}^{(1)}(z_A,z_A,\omega_{10})
=\frac{\omega_{10}e^{i\tilde{z}}}{2\pi c \tilde{z}^3}
\left(1-i\tilde{z}\right)
\end{equation}
[$\tilde{z}=2\omega_{10}(z_A-d)/c$]. Note that the off-diagonal
components of the Green tensor are equal to zero.
For \mbox{$z_A>d$}, the vdW potential then reads, in
accordance with Eqs.~(\ref{equ0})--(\ref{equ2}) and after neglecting
the off-resonant part,
\begin{align}
\label{equ20}
U(z_A)
&\simeq U_1^r(z_A)
\nonumber\\
&=-\mu_0\omega_{10}^2
\left[\operatorname{Re}G_{xx}^{(1)}
|\mathbf{d}_{10}^\parallel|^2
+ \operatorname{Re}G_{zz}^{(1)}
|\mathbf{d}_{10}^\perp|^2
\right]
\end{align}
[$\mathbf{d}_{10}^\parallel\! =\! ((d_{10})_x,(d_{10})_y,0)$,
$\mathbf{d}_{10}^\perp\! =\! (0,0,(d_{10})_z)$], where
\begin{multline}
\label{equ21}
\operatorname{Re}G_{xx}^{(1)}
=\frac{\omega_{10}}{4\pi c \tilde{z}^3}
 \left[\cos(\tilde{z})+\tilde{z}\sin(\tilde{z})\right.
-\left.\tilde{z}^2\cos(\tilde{z})\right]
\end{multline}
and
\begin{equation}
\label{equ21b}
\operatorname{Re}G_{zz}^{(1)}
=\frac{\omega_{10}}{2\pi c \tilde{z}^3}
 \left[\cos(\tilde{z})+\tilde{z}\sin(\tilde{z})\right].
\end{equation}
For $0<z_A\le d$, the potential is, as already mentioned, divergent
due to the unphysical assumption of vanishing absorption.
Hence the question may arise whether at least in the range $z_A>d$
the above-given result can be regarded as being
correct in the limit of sufficiently small absorption.
An answer to this question can obviously only be given
by taking material absorption into account from the very beginning,
as shall be done in the next section.


\subsection{Real (absorbing) LHM}

To allow for material absorption, we return to Eq.~(\ref{equ22})
for the scattering part of the Green tensor and set therein
$\varepsilon(\omega_{10})=-1+i\eta_e$, $\mu(\omega_{10})=-1+i\eta_m$, where
for simplicity we assume
$\eta\equiv\eta_e=\eta_m$.
Due to the positive imaginary parts of $\varepsilon$
and $\mu$, divergent integrals of the type of the second integral
in Eq.~(\ref{gtensor}) can never occur.
Note that nonvanishing absorption helps to
regularize the behavior of the integrals, just like it does in the
superlens geometry~\cite{Pendry2000}.

\begin{figure}[t]
\begin{center}
\includegraphics*[width=\linewidth]{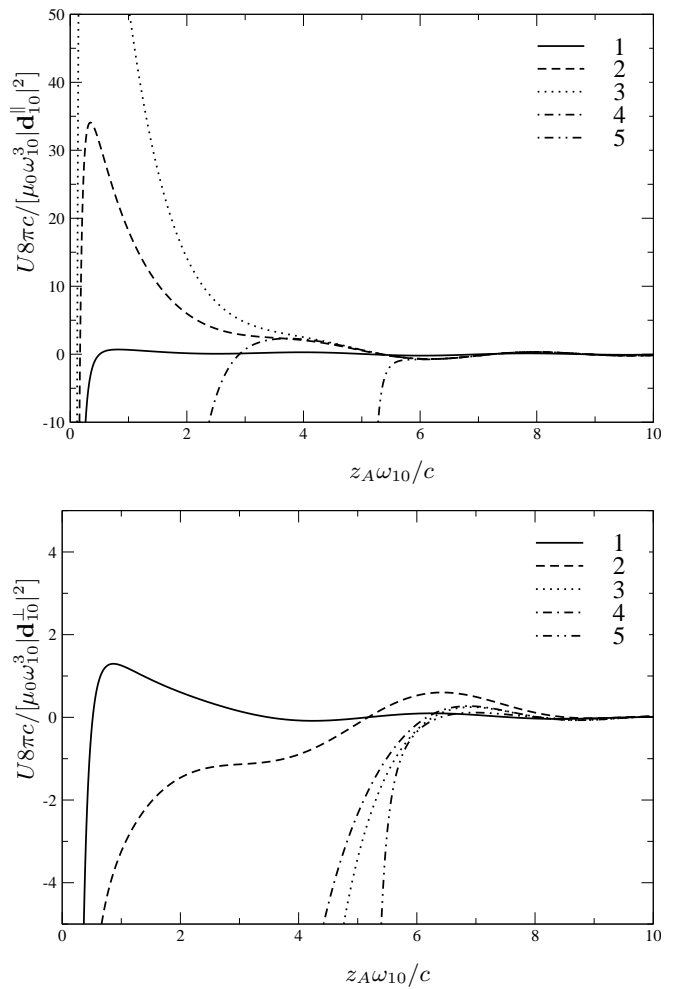}
\end{center}
\caption{Resonant VdW potential experienced by an excited
two-level atom in the setup sketched in Fig.~\ref{lhmfig} for
$d=5c/\omega_{10}$, $\varepsilon(\omega_{10})=\mu
(\omega_{10})=-1+i\eta$,
and (a) dipole moments parallel to the surface
and (b) dipole moments perpendicular to the surface.
Curves 1--5 refer to various degrees of absorption $\eta=10^{-1}$,
$10^{-3}$, $10^{-4}$, $10^{-5}$ and $0$, respectively. }
\label{fig2}
\end{figure}

In Fig.~\ref{fig2}, the resulting (resonant) vdW potential
is plotted versus the distance from the atom to the LHM slab,
for the two cases of parallel and perpendicular alignment of the
atomic dipole moment with respect to the slab and
for different values of absorption.
It is seen that the potential features an attractive behavior
in the nonretarded regime, which is governed by an inverse power law,
while in the retarded regime an oscillating
behavior with alternating sign of the potential occurs.
The nonretarded regime is dominantly determined by
evanescent waves [first integral in Eq.~(\ref{equ22.1})]
while propagating waves are mainly responsible for the
retarded regime [second integral in Eq.~(\ref{equ22.1})].
The figure further reveals that in
the limit of vanishing absorption when $\eta$ tends to zero, then
the potential approaches the one found, for $z_A > d$, in the case of
absolutely nonabsorbing LHM [Eq.~(\ref{equ20}) together with
Eqs.~(\ref{equ21}) and (\ref{equ21b})]. It is clearly seen
that in practice unavoidable absorption always prevents the
system to feature an infinitely negative potential in the interval
$0 < z_A \leq d$, as would be expected from the study of a nonabsorbing LHM.
We do, however, note that for very small absorption, the potential
starts to become strongly negative around the region $z_A\approx d$,
where this effect is more
noticeable on the case of perpendicular atomic dipole moment.

A very striking feature of the potential is completely missing
if absorption is disregarded:
For a transition dipole moment parallel to the surface
and medium absorption, a barrier appears
in front of the slab at distances $z_A\omega_{10}/c\lesssim 1$
[Fig.~\ref{fig2}(a), $10^{-4}\lesssim\eta\lesssim 10^{-3}$].
The fact that barriers occur only for transition dipole moments
parallel to the surface but not for those perpendicular to
the surface suggests that $s$-polarized waves, which are coupled
to the first but not the latter, play an important role in their
formation.

\begin{figure}[t]
\begin{center}
\includegraphics*[width=\linewidth]{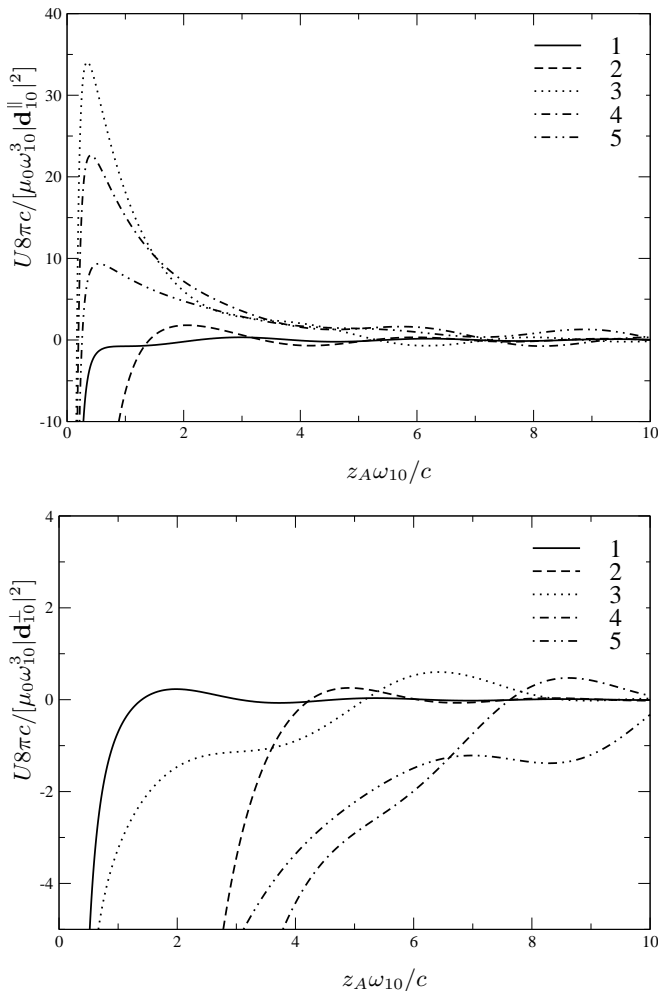}
\end{center}
\caption{
Resonant vdW potential experienced by an excited atom in
the setup sketched in Fig.~\ref{lhmfig} for
$\varepsilon(\omega_{10})=\mu
(\omega_{10})=-1+i 10^{-3}$,
and (a) dipole moments parallel to the surface
and (b) dipole moments perpendicular to the surface.
Curves 1--5 refer to various thicknesses of the slab $d\omega_{10}/c=0$,
$3$, $5$, $7$ and $10$, respectively.}
\label{fig3}
\end{figure}

For sufficiently weak absorption
[$\eta=10^{-5}$ in Fig.~\ref{fig2}(a) and
$\eta=10^{-3}, 10^{-4}, 10^{-5}$ in Fig.~\ref{fig2}(b)],
an attractive potential starts to appear at distances of a few
wavelengths, somewhat away from the surface. Atoms located within this
range will get adsorbed to the surface. This behavior is more
pronounced for a transition dipole moment perpendicular to the surface
than for a transition dipole moment parallel to the surface---a fact
that can be related to the coupling respective no coupling of these two
dipole moment orientations to $s$-polarized waves. By setting the
denominators in Eqs.~(\ref{r2-s}) and (\ref{r2-p}) equal to zero,
one finds that the poles of the generalized reflection coefficients
for $p$- and $s$-polarized waves, respectively, are solutions to the
equations
\begin{align}
\label{polea}
&   \frac{\beta_1}{\varepsilon
\beta} = \coth(i\beta_1 d),
\\
\label{poleb}
&   \frac{\beta_1}{\mu
\beta} = \tanh(i\beta_1 d).
\end{align}
These are the dispersion relations for the surface plasmon polaritons
\cite{Ruppin2001}, which can be strongly excited when the atom is
close to the surface. The real parts of the poles lie in the regions
covered by the second integral in Eq.~(\ref{equ22.1}).
The poles of the $s$- ($p$-)polarized waves are in the upper (lower)
half of the complex plane, where for smaller absorption
they are closer to the real axis.

The influence on the vdW potential of the slab thickness is
illustrated in Fig.~\ref{fig3}. Note that zero slab thickness means
an atom in front of a perfect mirror. As the slab thickness increases, a
potential barrier arises and grows in height for dipole moments
parallel to the surface. However, at some threshold value of $d$, the
height of the barrier starts to be reduced, and eventually the barrier
disappears when the slab is too thick. This can be explained as
resulting from the increasing effects of material absorption. It is
seen from the figures that the distance from the surface at which a
relatively strong attractive potential can occur increases with the slab
thickness. This is more pronounced for perpendicular
dipole moments rather than for parallel ones---a consequence of the fact that,
as already mentioned, the first is coupled more efficiently to the evanescent waves.


\subsection{Near-surface limit}
\label{sec3.3}

Further insight into how the appearance of the barrier depends on the
amount of absorption, the orientation of the atomic dipole moment as
well as the thickness of the LHM slab can be gained by examining
the vdW potential in the near-surface limit. Near the surface,
the evanescent waves dominate the potential
[the second integral in Eq.~(\ref{equ22.1})].
Let us consider the (resonant part of the)
potential as given by Eq.~(\ref{equ1}) together with
Eqs.~(\ref{equ22}) and (\ref{equ22.1})
for very short distances,
$z_A\omega_{10}/c\ll 1$, in more detail. The main contribution
to the $q$-integral is from values
$q\gg \omega_{10}/c$ and $q\gg\sqrt{|\varepsilon \mu|}\omega_{10}/c$
[$\varepsilon\equiv \varepsilon(\omega_{10})$,
$\mu\equiv \mu(\omega_{10})$],
since the exponential factor [in the second integral
in Eq.~(\ref{equ22.1})]
effectively limits the integration interval to
values $q\lesssim z_A$. Hence, with
\begin{equation}
\beta \simeq iq, \quad \beta_1\simeq iq
\end{equation}
the potential in the short-distance regime reads
\begin{multline}
\label{barrier}
U(z_A)=-\frac{\omega_{10}^2\mu_0}{8\pi}
\int _0^\infty \mathrm{d}q\;e^{-2qz_A}
\\
\times\! \left[\!\left(\operatorname{Re}r_{2-}^s
\!+\!\frac{q^2c^2}{\omega_{10}^2}
\operatorname{Re}r_{2-}^p\right)\!
|\mathbf{d}_{10}^\parallel|^2
\!+\!\frac{2q^2c^2}{\omega_{10}^2}
\operatorname{Re}r_{2-}^p\!
|\mathbf{d}_{10}^\perp|^2
\right]\!,
\end{multline}
where
\begin{multline}
\label{refl1}
  {\rm Re}\;r_{2-}^p =
  \left[ \left(|\varepsilon
|^2-1\right)\left(1+e^{-4qd}\right)
\right.
\\
\left.
  +\left(|\varepsilon
\!-\!1|^2+|\varepsilon
\!+\!1|^2\right)e^{-2qd}
  \right]
  \left|\varepsilon
\!+\!1\!+\!(\varepsilon
\!-\!1)
  e^{-2qd}\right|^{-2},
\end{multline}
\begin{multline}
\label{refl2}
   {\rm Re}\;r_{2-}^s\! =
   \left[ \left(|\mu
|^2-1\right)\left(1+e^{-4qd}\right)
\right.
\\
\left.
   - \left(|\mu
\!-\!1|^2\!+\!|\mu
\!+\!1|^2\right)e^{-2qd} \right]
     \left|\mu
\!+\!1\!-\!(\mu
\!-\!1)
     e^{-2qd}\right|^{-2}.
\end{multline}
When
\begin{equation}
\label{cond}
  \varepsilon
\simeq -1 \quad \mbox{and}\quad
  \mu
\simeq -1,
\end{equation}
then the first terms in Eqs.~(\ref{refl1}) and (\ref{refl2})
approximately vanish, thus
\begin{align}
\label{refl1a}
  &{\rm Re}\,r_{2-}^p  \simeq \frac
  { \left(|\varepsilon
  -1|^2+|\varepsilon+1|^2\right)e^{-2qd} }
  { \left|\varepsilon +1+(\varepsilon-1)
  e^{-2qd}\right|^2}\,,
\\
\label{refl2a}
   &{\rm Re}\,r_{2-}^s \simeq
   - \frac{ \left(|\mu -1|^2+|\mu +1|^2\right)e^{-2qd} }
     { \left|\mu +1-(\mu -1)
     e^{-2qd}\right|^2}\,,
\end{align}
where the opposite signs imply that the two polarizations give
competing contributions to the potential.
Namely, the $p$-polarized waves give rise to attractive contributions to
the potential while the $s$-polarized waves lead to repulsive ones.
Very close to the surface, due to the presence of the $q^2$ factor
[see Eq.~(\ref{barrier})], the contribution to the potential
of the $p$-polarized waves is proportional to $1/z_A^3$ while the
contribution of the $s$-polarized waves is proportional to $1/z_A$.
The contribution of the $p$-polarized waves hence dominates, resulting
in an attractive potential (also see Figs.~\ref{fig2} and~\ref{fig3}).
At some distance from  the surface, the contribution of the $s$-polarized
waves can dominate under appropriate conditions, which then leads to
the appearance of a potential barrier. This also explains the absence
of the barrier in the case where the dipole moment is perpendicular
to the surface.

Inspection of
Eqs.~(\ref{refl1})
and (\ref{refl2})
reveals that 
${\rm Re}\,r_{2-}^p$ is positive for all $|\varepsilon|\geq 1$ while ${\rm Re}\,r_{2-}^s$ is negative for all $|\mu|\geq 1$.
Since the height of the
potential barrier is determined by the magnitude of
$|{\rm Re}\,r_{2-}^s|$,
lefthandedness is not 
necessarily required to realize
a potential barrier
in general.
We note here yet another situation where, in the absence of the
lefthandedness property, a high potential barrier
might appear. For this it is convenient to consider
a more general setup
where the spatial area~2 in Fig.~\ref{lhmfig} is filled with
a medium of permittivity $\varepsilon_2(\omega)$
and permeability $\mu_2(\omega)$. Then, apart from a scaling factor due
to local-field correction,
$9\varepsilon_2^2(\omega_{10})/[2\varepsilon_2(\omega_{10})+1]^2$
\cite{Sambale2007},
the vdW potential retains the form of Eq.~(\ref{barrier}),
with the reflection coefficients~(\ref{refl1}) and (\ref{refl2}) being
modified as follows:
$|\varepsilon
|^2-1$ ($|\mu
|^2-1$) is replaced
by $|\varepsilon
|^2-|\varepsilon_2|^2$
($|\mu
|^2-|\mu_2|^2$) and
$\varepsilon
\pm 1$ ($\mu
\pm 1$) is replaced
by $\varepsilon
\pm \varepsilon_2$ ($\mu
\pm \mu_2$).
If we choose
$|\varepsilon
|> |\varepsilon_2| \quad \mbox{and}\quad |\mu
| < |\mu_2| $, the first terms, rather than the second terms, in the
modified equations~(\ref{refl1}) and (\ref{refl2})
dominate, but it remains that
$p$- ($s$-)polarized waves give rise to an attractive (repulsive) potential.
In this scenario, the appearance and the
height of
the barrier are determined by the difference between the
absolute values of the permittivity on the two sides of the interface
and the
accordant
difference between the absolute values of the permeability.
Negative (real parts of the) permittivities and/or
permeabilities are allowed but not a prerequisite. The presence of a
potential barrier in this case
is essentially a surface effect, the finite thickness of the slab
and the perfect mirror playing only secondary roles.

Let us return to the influence of the thickness of the LHM slab. As can be
inferred from Eq.~
(\ref{refl2}),
the magnitude of $|{\rm
Re}\,r_{2-}^s|$ is about 1 for $d\rightarrow 0$ (slab absent), and
is typically determined
by a $e^{2qd}$ term otherwise. 
Therefore the presence of the
slab is crucial for the appearance of a potential barrier.
When the slab is very thick, 
the influence of the mirror vanishes
$e^{- 2qd}\rightarrow 0$
[cf.~Eqs.~(\ref{refl1}) and (\ref{refl2})], 
and it is not difficult to verify that
Eq.~(\ref{barrier}) reproduces the result for
the resonant part of the potential of an excited atom in front of
an interface (see, e.g., Ref.~\cite{Buhmann2004}),
which, in the nonretarded limit reads as
\begin{equation}
\label{intf}
U(z_A)
=
-\frac{C}{z_A^3} \frac{|\varepsilon
|^2-1}{|\varepsilon
+1|^2}
\end{equation}
[$C=[|\textbf{d}_{10}^\parallel|^2
+2
|\textbf{d}_{10}^\perp|^2
]/(32\pi\varepsilon_0)$].
Clearly, when the slab is too thick, then
the atom only feels the vacuum--slab interface.
Note that a repulsive behavior of the potential is also possible
in the case of non-magnetic materials provided that $|\varepsilon |<1$.
It is particularly apparent that $|U(z_A)|$ cannot become infinity
as $z_A\rightarrow 0$, because the
underlying macroscopic quantum electrodynamics is valid only
for atom--surface distances much larger than the average distances
between the constituents of the media involved.

Although potential barriers in planar magnetodielectric multilayer
structures have also been found in the case of ground-state atoms
\cite{Kampf2005}, it is worth noting that the barriers are generally
much more significant in the case of excited atoms. For instance, the
peaks of the potentials in the numerical examples given here are
about $6$ orders of magnitude larger than those given in
Ref.~\cite{Kampf2005}. This is because the potential in the case of
ground-state atoms is an integral over the whole frequency range, for
most of which the material does not exhibit strong electric or
magnetic responses. The excited atoms essentially allows one to pick out
a frequency of choice.

\section{Spontaneous decay rate revisited}
\label{spe}

Whereas the (resonant) vdW potential depends on the real part of the 
Green tensor, the rate of spontaneous decay is determined by its
imaginary part. From Eq.~(\ref{gtensor}) for the Green tensor of
the absolutely nonabsorbing setup considered in Sec.~\ref{sec3.1},
it can be seen that the contributions from
evanescent waves, which give rise to divergences in the region
$z_A\le d$, are purely real and thus do not contribute to the decay rate.
The decay rate is thus expressed in terms of traveling-wave contributions
which are finite for arbitrary $z_A$. However, baring in mind
the findings for the vdW potential, one should carefully examine
whether these idealized results valid for nonabsorbing material are an
appropriate approximation to the reality.

To do so, we note that
from Eq.~(\ref{gamma}), the decay rate can be rewritten as
\begin{multline}
\label{drate}
\frac{\Gamma}{\Gamma_0} \!=\! 1+ \frac{6\pi c}{\omega_{10}
|\mathbf{d}_{10}|^2}
\operatorname{Im}
\left[G^{(1)}_{xx}
|\mathbf{d}_{10}^\parallel|^2
+G^{(1)}_{zz}
|\mathbf{d}_{10}^\perp|^2
\right],
\end{multline}
where
\begin{equation}
\label{Gamma0}
\Gamma_0=\frac{
|\mathbf{d}_{10}|^2\omega_{10}^3}{3\pi\hbar\varepsilon_0c^3}
\end{equation}
is the decay rate of the atom in free space.
Let us first assume an absolutely nonabsorbing LHM having
\mbox{$\varepsilon=\mu=-1$} again. From Eqs.~(\ref{gr1}) and (\ref{gr2})
we obtain
\begin{align}
\label{imGx}
&\operatorname{Im}G^{(1)}_{xx}=\frac{\omega_{10}}{4\pi c\tilde{z}^3}
\left[\sin(\tilde{z})
-\tilde{z}\cos(\tilde{z})-\tilde{z}^2\sin(\tilde{z})\right],
\\[1ex]
\label{imGz}
&\operatorname{Im}G^{(1)}_{zz}
=\frac{\omega_{10}}{2\pi c\tilde{z}^3}\left[\sin(\tilde{z})
-\tilde{z}\cos(\tilde{z})\right]
\end{align}
[$\tilde{z}=2\omega_{10}(z_A-d)/c$].
Equation (\ref{drate}) together with Eqs.~(\ref{imGx}) and (\ref{imGz}),
which mathematically holds for any atom--surface distance,
is studied in Ref.~\cite{Fleischhauer2005}. It is not difficult to see that
$\operatorname{Im}G^{(1)}_{xx}$ and $\operatorname{Im}G^{(1)}_{zz}$
are even functions of $\tilde{z}$, and are finite at the surface.
The position $z_A=d$, termed focus point in
Ref.~\cite{Fleischhauer2005}, stands out in that the optical path
for a ray of light for a round trip to the mirror and back is zero.
For $z_A=d$, Eq.~(\ref{drate}) together with Eqs.~(\ref{imGx})
and (\ref{imGz}) implies complete inhibition of
spontaneous decay ($\Gamma\!=\!0$) for a dipole moment oriented parallel to the surface and
enhancement of spontaneous decay
($\Gamma\!=\!2\Gamma_0$) for dipole
moment oriented perpendicular to the surface~\cite{Fleischhauer2005}.

\begin{figure}[!t!]
\begin{center}
\includegraphics*[width=\linewidth]{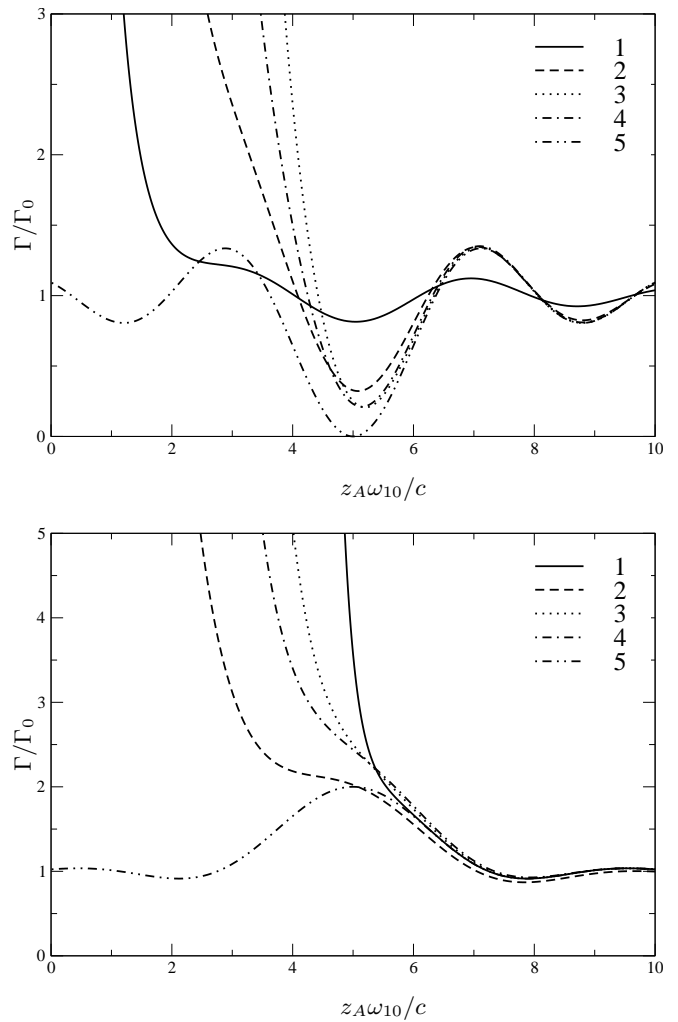}
\end{center}
\caption{Atom--surface distance dependence of the
decay rate of an excited atom in
the setup in Fig.~\ref{lhmfig} for
$\varepsilon
=\mu=-1+i\eta$,
$d=5c/\omega_{10}$,
and (a) dipole moments parallel to the surface
and (b) dipole moments perpendicular to the surface.
Curves 1--5 refer to various degrees of absorption $\eta=10^{-1}$,
$10^{-3}$, $10^{-4}$, $10^{-5}$ and $0$, respectively.}
\label{fig4}
\end{figure}

It is worth noting that absorption can drastically change this behavior.
Inclusion in the calculations of the effect of material absorption requires the
calculations to be performed on the basis of the exact (scattering part
of the) Green tensor as given in Eq.~(\ref{equ22}) together
with Eqs.~(\ref{r2-s})--(\ref{r21}).
Numerical examples are given in Fig.~\ref{fig4}, where the case of
zero absorption in accordance with Eqs.~(\ref{drate})--(\ref{imGz}) is
also shown to facilitate comparison.
We see that, whereas in the case of strictly zero absorption
the decay rate as a function of the atomic position $z_A>0$
is symmetric with respect to the position $z_A\!=\!d$,
any absorption destroys this symmetry. As a result, large
enhancement of the spontaneous decay can be observed when the
atom is near the surface, which is obviously due to the absorption-assisted
atomic coupling with evanescent waves---an effect already known
for ordinary materials (see, e.g., Ref.~\cite{Vogel2006})
which features qualitatively new distance terms
[cf.~Eq.~(\ref{drate2})].
This result is consistent with those reported in Ref.~\cite{Xu07}.
In particular for distances $z_A\leq d$,
Eq.~(\ref{drate}) together with Eqs.~(\ref{imGx}) and~(\ref{imGz})
can not be regarded as an acceptable approximation
to the spontaneous decay rate in
the case of small absorption---a result which
corresponds,  
for $z_A < d$,
to the result found in the case of the vdW potential.

\begin{figure}[!t!]
\begin{center}
\includegraphics*[width=\linewidth]{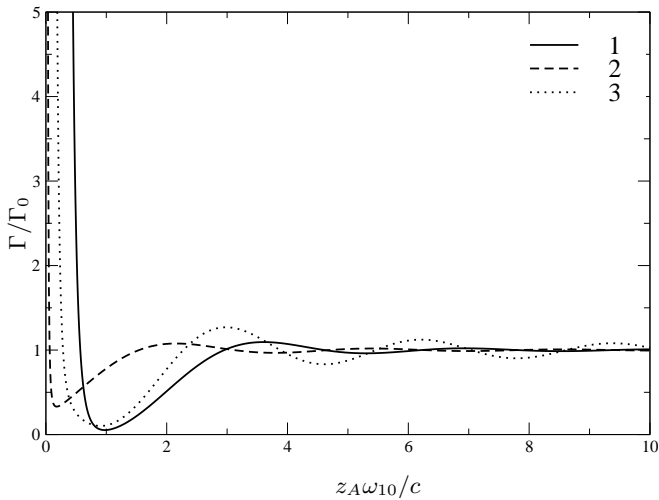}%
\end{center}
\caption{Atom--surface distance dependence of the
decay rate of an excited atom in the setup in
Fig.~\ref{lhmfig} for
$d\!=\!5c/\omega_{10}$, and (1)
$\varepsilon
\!=\!1$, $\mu
\!=\!-1+i 10^{-3}$, and
atomic dipole moment perpendicular to the surface;
(2)
\mbox{$\varepsilon
\!=\!-1+i  10^{-3}$}, $\mu
=1$, and
atomic dipole moment perpendicular to the surface;
(3)
$\varepsilon
\!=\!1$, $\mu
\!=\!-1+i 10^{-3}$, and
atomic dipole moment parallel to the surface.}
\label{fig5}
\end{figure}%

To further elucidate the influence of the evanescent waves, let us
examine the near-surface limit of the rate of spontaneous decay.
By using approximations similar to those
in Sec.~\ref{sec3.3},
it can be shown that for $z_A\omega_{10}/c\ll 1$
\begin{multline}
\label{drate1}
\frac{\Gamma}{\Gamma_0} = 1 + \frac{3c}{4\omega_{10}
|\mathbf{d}_{10}|^2
}
\int _0^\infty \mathrm{d}q\;e^{-2qz_A}
\Bigl[\Bigl(\operatorname{Im} r_{2-}^s
\\
+\frac{q^2c^2}{\omega_{10}^2}\operatorname{Im}
r_{2-}^p\Bigr)
|\mathbf{d}_{10}^\parallel|^2
+\frac{2q^2c^2}{\omega_{10}^2}\operatorname{Im}r_{2-}^p
|\mathbf{d}_{10}^\perp|^2
\Bigr],
\end{multline}
where
\begin{align}
\label{reflgamma}
&\operatorname{Im}r_{2-}^s=\frac{2{\rm Im}\mu
(1-e^{-4qd})}{|\mu
+1-(\mu
-1)e^{-2qd}|^2},
\\[1ex]
\label{reflgamma2}
&\operatorname{Im}r_{2-}^p
=\frac{2\operatorname{Im}\varepsilon
\left(1-e^{-4qd}\right)}
{|\varepsilon
+1-(\varepsilon
-1)e^{-2qd}|^2}.
\end{align}
Note that in this case only evanescent waves contribute to the rate.
Unlike the real parts
[cf. Eqs.~(\ref{refl1}) and (\ref{refl2})], the imaginary parts
of $r_{2-}^s$ and $r_{2-}^p$ have the same (positive) sign.
The two polarizations therefore
contribute constructively to the spontaneous decay rate.
If the slab becomes sufficiently thick, Eq.~(\ref{drate1}) reduces,
in leading order, to
\begin{equation}
\label{drate2}
\frac{\Gamma}{\Gamma_0}=1
+\frac{D}{z_A^3}\frac{\operatorname{Im}\varepsilon
}
{|\varepsilon
+1|^2}
\end{equation}
[$D=3c^3[
|\mathbf{d}_{10}^\parallel|^2
+2
|\mathbf{d}_{10}^\perp|^2]/(8\omega_{10}^3
|\mathbf{d}_{10}|^2
)$
], i.e., the spontaneous rate for an atom in front of an interface.
Equation~(\ref{drate2})
shows that the
decay rate takes on large values as $z_A\rightarrow 0$, which is a
consequence of direct energy transfer from the atom to the
constituents of the medium (see, e.g., Ref.~\cite{Vogel2006}).

It is well known that lefthandedness of the slab is of course
not a necessary condition for, say, inhibition of
spontaneous decay, as illustrated in Fig.~\ref{fig5}.
The figure shows that almost complete inhibition of spontaneous decay
is possible for both parallel and perpendicular orientations of the
dipole moment with respect to the surface of a slab that is not lefthanded.


\section{Concluding remarks and summary}
\label{sum}

The macroscopic theory employed in this paper breaks down on length
scales which are comparable to the size of the elementary building
blocks of the metamaterial employed. In order to observe the
predicted effects, one thus has to ensure that both the atomic
transition wavelength and the atom--surface separation are larger
than this length scale. With currently available metamaterials, this
can most readily be achieved with polar molecules whose rotational and
vibrational transition wavelengths can be very large. For example, the
bulk meta-material reported in Ref.~\cite{Paul2008} exhibits negative
refraction with $\operatorname{Re}n\approx -1.6$ at a frequency of
1.03 THz; it consists of unit cells with lattice constant $a\approx
6\times 10^{-5}\mathrm{m}$. The frequency of the first
rotational
transition of a NH molecule (about 1 THz \cite{Buhmann2008}) falls
within this frequency window, with the associated wavelength
%
being sufficiently large in comparison with the lattice constant. One
could expect the macroscopic results to be valid down to
molecule--surface separations of $z_A=10^{-4}\mathrm{m}$, which
corresponds to $z_A\omega_{nk}/c\approx 2$. Smaller values of
$z_A\omega_{nk}/c$ can be reached by using molecules with larger
transition wavelengths. 

Let us briefly comment on the strength of the force
$F(z_A)\!=\!- \partial U(z_A)/\partial z_A$
associated with a potential of the type shown
in Figs.~\ref{fig2} and \ref{fig3}.
In this context, it may be convenient to express
the potential and the force
in terms of the spontaneous decay rate~$\Gamma$.
From Eqs.~(\ref{equ0}), (\ref{equ1}), and~(\ref{Gamma0})
it follows that $U(z_A)\!=\!-3/2\hbar \Gamma_0
\lambda_{10}{\rm Re}\, G^{(1)}_{xx}(z_A,z_A,\omega_{10})$,
where a dipole moment parallel to the surface has been assumed.
Consider an excited hydrogen-like atom with a
transition having spontaneous decay rate $\Gamma_0 \!\sim\! 6
\times 10^8 \,\mathrm{s}^{-1}$ and wavelength
$\lambda_{10}\! \sim\!
10^{-7}\,\mathrm{m}$ \cite{Nist}. If the atom is at
rest relative to the slab, the force that can counter the gravitation
force and so levitate the atom has to be larger than $mg \!=\!
3\times10^{-26}\,\mathrm{N}$. This can be achieved easily. For example, the peak
value of the force corresponding to the potential barrier represented
by the dotted line in Fig.~\ref{fig2}(a) is $10^{-16}\,\mathrm{N}$.

Consider now a situation where one has a sample of a
dilute gas of excited atoms at some temperature. Take again
hydrogen-like atoms with the same excited level as above.
At a temperature of $T\!=\!10\,\mathrm{K}$,
the kinetic energy of each particle
is $E\!=\!3kT/2\!\sim\! 2\times 10^{-22}\,\mathrm{J}$.
A LHM slab of thickness
$d\omega_{10}/c\!=\!9$, permittivity and permeability
$\varepsilon
\!=\!\mu\!=\!-1+i 2\times 10^{-6}$
then provides a potential barrier of a height of
$U\!=\!3\times10^{-22}\,\mathrm{J}$. This
potential barrier is high enough to levitate the sample of
the dilute atomic gas.
With an appropriate arrangement, it clearly
can be employed as a trapping mechanism. One may also think of
an (evaporative) cooling apparatus. The kinetic energy is an averaged
quantity. Some atoms among the gas sample can have kinetic energy
exceeding the average and higher than the potential barrier. These can
overcome the barrier and get adsorbed by the walls, in so doing
lowering the average kinetic energy of the remaining (a little more
diluted) gas sample.
It should be pointed out that in the examples
meta-materials with very small absorption
are assumed, which are still beyond the reach of
today's experimental techniques.
Furthermore, since excited atoms are considered, the
effects predicted hold only for
time periods short compared to the atomic lifetime. A more refined
theory should include the broadening and shifting of the atomic
transition frequency.

In summary, we have studied the
resonant vdW potential experienced by an
excited two-level atom in front of a meta-material slab backed by a
perfect mirror, with emphasis on
a LHM slab with unity negative real part of both
the permittivity and the permeability,
and an atom located in vacuum. We have shown that
material absorption has to be taken into account
in order to avoid divergent values of the potential
when the atom--surface separation becomes
smaller than the slab thickness.
Only for values of the atom--surface separation which are larger
than the slab thickness,
disregarding of absorption leads to reasonable results provided
that absorption is sufficiently small.

Further, we have shown that the setup can be used to
form potential barriers for excited atoms -- barriers that
are typically of several orders of magnitude higher
than those observed for ground-state atoms. In particular, they are
significant enough to suggest potential use in quantum levitation,
trapping, or cooling of atoms.
The barriers are results of competing
contributions from $s$- and $p$-polarized waves, thus are peculiar
to atoms with dipole moments parallel to the surface.
For random atomic dipole moment orientations, a corresponding
averaging has to be applied
which may lead to a reduction of the barrier height.

Finally, we have revisited the
effect of the LHM slab on the spontaneous decay
of the excited atom. We have shown that already
an arbitrarily small but finite amount of material absorption
can drastically change the decay compared to that predicted when
absorption is ignored. In particular near the surface, absorption gives rise
to a noticeable enhancement of the decay.
As a prospective task, the theory
could be expanded to other systems such as spherically or
cylindrically symmetric ones, amplifying instead of absorbing media
could be assumed, and the vdW interaction between excited atoms near a
macroscopic body of meta-material could be investigated.


\acknowledgments

The work was supported by Deutsche Forschungsgemeinschaft.
We acknowledge funding from the Alexander von Humboldt
Foundation (H.T.D. and S.Y.B.)
and the Vietnamese Basic Research Program (H.T.D.).

\end{document}